# Optimal morphometric factors responsible for enhanced gas exchange in fish gills

Prasoon Kumar[a,b,c], Prasanna S Gandhi[b], Mainak Majumder[c]

a IITB-Monash Research Academy, Powai, Mumbai, Maharashtra-400076, India

b Suman Mashruwala Advanced Microengineering Laboratory, Department of Mechanical Engineering, Indian Institute of Technology Bombay, Powai, Mumbai, Maharashtra 400076, India

c Nanoscale Science and Engineering Laboratory (NSEL), Department of Mechanical and Aerospace Engineering, Monash University, Clayton, Melbourne, Australia

## Abstract

Fish gills are one of the most primitive gas/solute exchange organs, having the highest ventilation volume, present in nature. Such performance is attributed to a functional unit of gill - secondary lamella - that can extract oxygen from an ambience even at a very low partial pressure. For centuries, gills have stood as one of the simplest but an elegant gas/solute exchange organs. Although the role of various morphometric factors of fish gills on gas/solute exchange capabilities have been reported, there has been limited understanding on what makes fish gills as an excellent gas/solute exchange system. Therefore, in the current study, we have theoretically studied the variation of few structural and parametric ratios, which were known to have role in gas/solute exchange, with respect to the weight of fishes. Thereafter, modelling and simulation of convection-diffusion transport through a two dimensional model of secondary lamella were carried out to study different factors affecting the performance of gills. The results obtained from both the studies (theoretical and computational) were in good agreement with each other. Thus, our study suggested that fish gills have optimized parametric ratios, at multiple length scales, throughout an evolution to arrive at an organ with enhanced mass transport capabilities. Further, our study also highlighted the role of length of primary and secondary lamella, surface area of secondary lamellae and inter-lamellar distance on gas/solute exchange capabilities of fish gills. Thus, these defined morphological parameters and parametric ratios could be exploited in future to design and develop efficient gas/solute exchange microdevices.

# 1. Introduction

The separation of gases, solutes and solvents under convective flow condition is a requirement in number of applications like artificial kidney dialyser, extra corporeal membrane separators, water filtration, desalination and others [1, 2]. The biomimcry has been used to solve above engineering problems in the recent years by drawing an inspiration from natural systems like avian lung, human lungs, human kidney, human vasculature etc. [3-5]. However, the challenges of developing compact, efficient gas/solute exchange devices still persist. To achieve the above objective, the design of a device need to be improved by taking inspiration from a simple, yet elegant, gas/solute separating organ system. One of such primitive known gas/solute separating organ systems in nature is fish gills. Gills can be taken as a model design for the development of better gas/solute exchanger due to their ability to extract oxygen from water even at a very low partial pressure, having the highest ventilation volume among all species, having design leading to continuous efflux of oxygen from ambient to the blood vessels and an overall compactness of the organ system [6]. Therefore, fish gill needs to be thoroughly studied at different length scales to determine the reasons for it to be an excellent gas/solute exchange organ in nature for centuries. The knowledge of fish gills' design generated through above study can lead to development of microfluidic devices for better gas/solute separation.

The gills from different varieties of fishes have been studied for a long time from their physical, morphological, structural and biochemical point of view[7-11]. However, the knowledge of fish gill architecture and morphometrics in the light of their gas/solute exchange capabilities still need considerable attention. The basics structures of gills and their role in gas/solute exchange have been well elucidated in detail through a series of publications by Hughes et al. Through anatomical studies, they have calculated the dimensions of gills of different fishes at multiple length scales. Further, they have also established the role of above determined morphometric factors like thickness of blood-water barrier, size and arrangement of secondary lamellae, on the gas exchange capacities in different weights of fishes[10, 12]. Their studies suggested that surface area of gills, which is responsible for gas/solute exchange, scales with the weight of fishes. Another prominent finding which they reported was the counter-current flow of blood and water stream in secondary lamella and inter-lamellar space, respectively. Later, theoretical model was developed by J. Piiper and P. Scheid to study the gas diffusion in different organ systems.

Their model suggested that the effect of counter current flow on gas diffusion is better than the cross-current flow and uniform pool system, assuming the same ventilation perfusion conductance ratio across an epithelial layer. They further substantiated their results through experimental findings in a fish, a bird and a mammal where above three modes of gas exchange operated. In addition, they also commented that the enhanced gas exchange in counter-current flow occurs in a limited range of medium-to-blood conductance ratio (around 1)[13, 14]. Their model beautifully captured the role of counter-current flow on gas diffusion effectiveness, but the role of dimensions of secondary lamella of fish gills on effectiveness of gas/solute exchange remained elusive, except for an enhanced surface area. Later, studies carried out by K Park et al demonstrated an existence of optimal lamellar distances in fish gills irrespective of their mass for an enhanced gas exchange. Their theory and experiments effectively captured the poor dependence of inter-lamellar distance on weight of fishes. Thus, they established the evolutionary conservation of these inter lamellar distances in fishes[15].

Although, above studies identified the role of different morphometric factors on gas/solute exchange, seldom has the study been done to the best of our knowledge to study the structural organization of gills as an organ and parameterization of these morphometric factors in fishes. Thus, there is a need to study and explore dimensional features of a secondary lamella and its functional relationship with gas/solute transport in fishes. Therefore, in our current work, we have theoretically studied the variation in different morphometric features and parametric ratios of fish gills in different sizes of fishes form data reported in literature. Thereafter, computationally investigation of the convective-diffusion process in a simplified model of secondary lamella was carried out to illustrate the role of parametric ratios on gas/solute exchange in a fish gills. Our data analysis suggested that the values of these parametric ratios remain nearly conserved in fishes with different body mass and find good agreement with the values predicted by our computational study. Thus, the study enabled a better understanding of functioning of fish gills. The work also laid a foundation for the design parameters worth attention, before developing a bio-inspired fish's gill devices for application in heat and mass exchanger.

## 2. Structure and function of fish gills

One of the most efficient gas/solute exchange systems developed in an early evolutionary stage of living beings were that of gills in fishes. The outstanding functioning of fish gills as an excellent gas exchanger is without any trade-offs. The structure of fish gills support the transfer of gases and metabolites efficiently from an aquatic surrounding to a fish's body and vice versa. Moreover, the fish gills are capable of extracting oxygen from the water medium under extreme environment; from brackish sea water to murky lakes where oxygen partial pressure are abysmally low[16, 17].

The uniqueness of a fish gill lies in its hierarchical multi-scale structure and functioning. The fish gills' hierarchical structure is shown in the Figure 1. The fish gills are the first vertebrate gas exchange organ, which comprises parallel array of thin epithelial matrix encasing complex bifurcating vasculature known as secondary lamella. These secondary lamellae are stacked in parallel over a primary lamella which in itself is arranged in a rack like structure as shown in the Figure 1. Such arrangements of functional unit of fish gills dramatically increase the surface area to volume ratio. The thin epithelium layer in secondary lamellae provides a shuttle barrier between an aquatic environment and a fish's blood running in vascular network encased in secondary lamella is shown in the Figure 1. The primary lamella is attached to a stem where two primary blood vessels, namely; efferent artery and vein run in parallel. They are further subdivided into capillaries and sub-capillaries while entering into secondary lamellae. When fishes swim in water, the fishes gulp water through their mouth and direct them to flow through inter-lamellar space. The concentration gradient of gases/solute across an aquatic environment and fish's blood drives an exchange of gas and metabolites through high surface area offered by thin epithelial barrier of secondary lamellae. Meanwhile, they also pump blood through blood capillaries encased in secondary lamella with a velocity enough to allow maximum diffusion of gas/solute from water flowing in inter lamellar spaces. Thus, the presence of fractal design of micro/nano capillaries encased in thin secondary lamella supports efficient flow of blood. They also execute majority of functions like respiration, excretion single headedly, which are carried out by pulmonary and renal system in mammals [7]. The exchange of gases/solutes is primarily driven by convection-diffusion phenomena that occur at a site of each secondary lamella of fish gills [9].

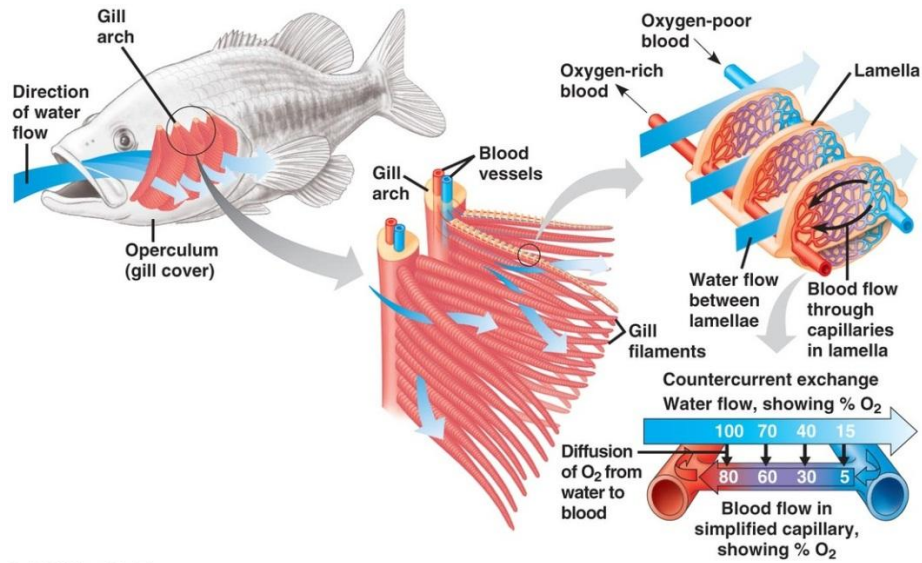

Figure 1 Schematic of multi scale architecture of fish gill demonstrating exchange of oxygen from ambient water to fish body[21]

The water flows through inter lamellar spaces between secondary lamellae of fish gills. The cross-current flow of water and blood in between secondary lamellae and blood capillaries respectively, and corresponding concentration gradient of gases and metabolites causing highly efficient diffusion pathway, produces high mass transfer ratio. This counter-current flow system always maintain blood with low oxygen concentration as compared to oxygen concentration of water flowing between the secondary lamellae of fish gills; and consequently continual diffusion of oxygen into the blood. The schematic in the Figure 1 explains role of cross current flow system in maximum possible gas and metabolites exchange between external aquatic environment and internal blood vessels [11]. This clearly illustrates the role of fluid flow direction in fish gills to maintain sustained concentration gradient and continuous diffusion flux of gases/solutes.

## 3. Materials and Methods

### 3.1 Theoretical study of fish gills

The morphometric data of fish gills were obtained from the work of Hughes et.al[11] shown in the Table 1.

**Table 1 Morphometric features of fish gills**

|  | Wt of fish (gm) | Inter-lamellar dist (mm) | Avg. PL length (mm) | Number of SL/mm | Length of SL (mm) | Hgt. of SL (mm) |
|---|---|---|---|---|---|---|
| *Trachurus trachurus* | 12 | 0.017 | 2.37 | 38 | 0.18 | 0.07 |
|  | 40 | 0.017 | 4.18 | 39 | 0.25 | 0.1 |
|  | 125 | 0.017 | 6.16 | 38 | 0.7 | 0.2 |
|  | 135 | 0.02 | 6.14 | 39 | 0.75 | 0.15 |
| *Clupea harengus* | 85 | 0.018 | 4.46 | 33 | 0.6 | 0.17 |
| *Gadus merlangus* | 51 | 0.03 | 3.18 | 21 | 0.6 | 0.2 |
| *Onos mustela* | 20 | 0.023 | 2.39 | 26 | 0.45 | 0.15 |
|  | 80 | 0.03 | 4.66 | 20 | 0.55 | 0.27 |
| *Crenilabrus melops* | 65 | 0.04 | 2.96 | 21 | 0.85 | 0.3 |
| *Salmo trutta sp.* | 175 | 0.023 | 5.44 | 21 | 0.7 | 0.2 |
| *Tinca tinca* | 140 | 0.025 | 8 | 25 | 0.86 | 0.1 |
| *Chaenocephalus sp.* | 750 | 0.053 | 11.96 | 12 | 1.6 | 0.5 |
|  | 790 | 0.057 | 9.84 | 12 | 1.5 | 0.55 |
| *Lophius piscatorius* | 1550 | 0.07 | 13.31 | 11 | 1.1 | 0.4 |
| *Plmronectes platessa* | 86 | 0.04 | 5.08 | 20 | 0.75 | 0.2 |
| *Zeus faber* | 300 | 0.04 | 4.97 | 15 | 1.1 | 0.11 |
| *Trigla gurnardus* | 18 | 0.02 | 1.84 | 22 | 0.35 | 0.06 |
| *Cottus bubalis* | 40 | 0.04 | 2.86 | 16 | 0.8 | 0.3 |
|  | 52 | 0.04 | 3 | 16 | 0.8 | 0.25 |
| *Callionymus lyra* | 64 | 0.05 | 2.86 | 15 | 0.8 | 0.3 |
|  | 46 | 0.03 | 2.56 | 16 | 0.7 | 0.25 |
|  | 24 | 0.047 | 1.77 | 17 | 0.6 | 0.15 |

Thereafter, the dimensions of different features of fish gills were analyzed for their variation with respect to the weight of different fishes and reported as graphs after considering below dimensionless parameters.

a) Thickness_mem/width_water - the ratio width of epithelium and water channels ($H_2/H_1$)

b) Width_bld/width_water - the ratio width of blood and water channels ($H_3/H_1$)

c) Blood_parameter - the ratio of the diffusivity and speed in the blood layer ($D_3 \times L$)/ ($V_3 \times H_1^2$)

d) Water_parameter - the ratio of the diffusivity and speed in the water layer (D1xL)/(V1xH1^2).

For the purpose of the calculation of these dimensional less parameters for different fishes, following assumptions were considered.

1) The diameter of a single micro capillary is the maximum width of blood channel spanning across the width of the secondary lamella.
2) The flow calculations were done assuming the rectangular cross-section of inter lamellar space and secondary lamella.
3) The tortuosity and branching of blood channel and its diameter variation along the length are neglected for simplifying the model.
4) The enzymatic reaction based capture of oxygen by hemoglobin in the blood was neglected and only diffusion was considered.

### 3.2. Computational study of convective- diffusion phenomenon in fish gills

To model and simulate the convection diffusion phenomenon in a secondary lamellae, we have studied the phenomenon on a two dimensional model of a secondary lamella as shown in the Figure 2.

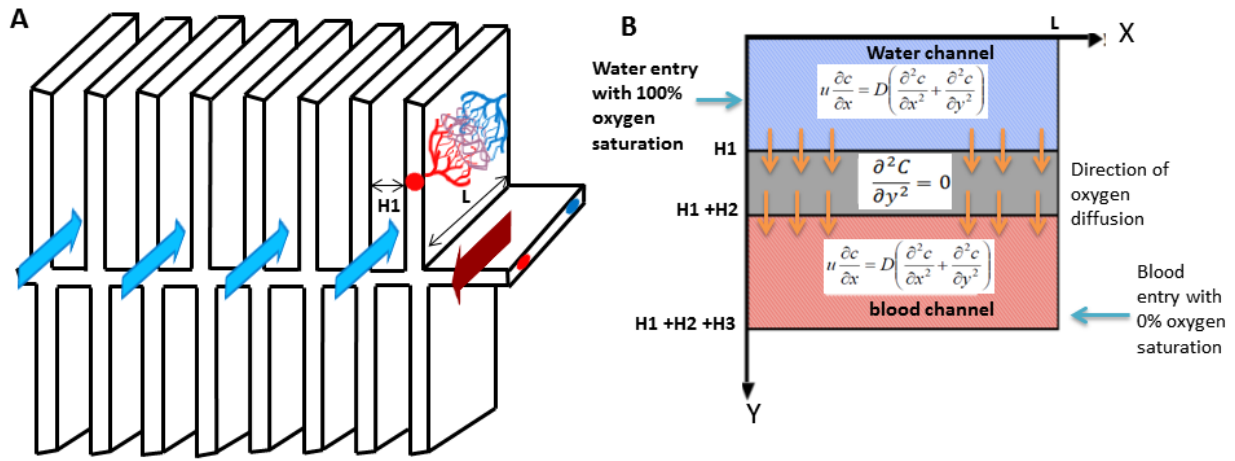

Figure 2 A) A schematic of secondary lamellae stacked in parallel on primary lamella. The blue arrow indicates the direction of water and red arrow represents direction of blood B) A 2D model of secondary lamella used in computational simulation

We have made certain assumptions while simulating the diffusion phenomenon in secondary lamella. These assumptions are

1) A two dimensional model of gas exchange in a single secondary lamella is considered
2) Blood channel represents the width of blood vessel inside the tissue matrix of secondary lamella
3) Water channel width is the width of the inter lamellar space
4) The gray colored domain between blood channel and water channel represents the thin epithelial matrix, a barrier separating blood and water.
5) Steady state convective-diffusion phenomenon is solved in above model
6) The blood and water flow velocity is assumed to be average velocity rather than a velocity with a parabolic profile and flowing in counter-current direction
7) Diffusion of oxygen is considered from water channel to blood channel by crossing thin epithelial matrix.

Under above assumptions convective- diffusion phenomenon was modeled in COMSOL Multiphysics 4.3b. The governing equations in each domain are

$$U_w \frac{\partial C}{\partial x} = D_w \left( \frac{\partial^2 C}{\partial x^2} + \frac{\partial^2 C}{\partial y^2} \right) \ldots \ldots \text{(Water domain)}$$

$$U_b \frac{\partial C}{\partial x} = D_b \left( \frac{\partial^2 C}{\partial x^2} + \frac{\partial^2 C}{\partial y^2} \right) \ldots \ldots \text{(Blood domain)}$$

$$D_e \frac{\partial^2 C}{\partial y^2} = 0 \ldots \ldots \text{(Epithelial barrier)}$$

$$N = -D \frac{\partial C}{\partial x} + U_x C$$

Where, $U_w$ is average velocity in water channel, $U_b$ is average velocity in blood domain, $D_w$ is the diffusion coefficient of oxygen in water, $D_b$ is the diffusion coefficient of oxygen in Blood, $D_e$ is the diffusion coefficient of oxygen in epithelial matrix, C is the concentration of oxygen and N is the flux

Initial conditions : C (x,y) = 0

Boundary condition: inlet at water channel → C = $C_o$

Outlet at water channel → n.N = 0

Inlet at blood channel → C = 0

Outlet at Blood channel → n.N = 0

**Table 2 Parameters used for modelling of convective diffusion phenomenon in secondary lamella**

| Parameters | Value | Parameters | Value |
|---|---|---|---|
| Width of water channel | 35μm | Diffusion coefficient of oxygen in Blood | 2.18e-9m$^2$/sec |
| Width of blood channel | 8.8μm | Diffusion coefficient of oxygen in water | 3.4e-9m$^2$/sec |
| Thickness of epithelial layer | 2.8μm | Diffusion coefficient of oxygen in epithelial layer | 1.1e-9m$^2$/sec |
| Length of channels | 760μm | Density and viscosity of water | 1000kg/m$^3$ 0.001kg/ms |
| Avg. velocity of water | 0.014m/s | Density and viscosity of blood | 1050kg/m$^3$ 0.0035/kg/ms |
| Avg. velocity of blood | 0.0005m/s | Conc. Of Oxygen in water stream | 0.25moles/m$^3$ |

Thereafter, considering the initial and boundary condition with the values of parameters in the Table 2 [11, 18, 19], we have studied the concentration of oxygen at the exit of blood channel by varying the different parametric ratios under steady state condition.

## 4. Results and discussion

In order to explore the role of primary and secondary lamellae, we analyzed the length of primary and secondary lamella with respect to fish's body masses. We observed that length of secondary lamellae and primary lamellae followed a logarithmic relation with fish body weight Figure 3. However, the length dependence of primary lamellae on a fish body mass was more pronounced as compared to secondary lamella. This might be primarily because secondary lamella being a site of solute/gas exchange, its length is perhaps guided by diffusion length of solute/gas during a convective flow. Due to evolutionary conservation of the inter-lamellar distance across different weight of fishes [15], the time scale of diffusion of gases/solutes across the inter-lamellar space should be less than the residence time of water flowing in the inter-lamellar channel. However, the difference between the two time scales should not be too high and residence time of water in a convective flow should be only marginally higher than diffusion time scale. Moreover, considering the pressure drop along the length is confined by pumping capacity of fishes, the variation in the flow velocity in an lamellar space is minimal and remain laminar [11, 15, 20]. Therefore, length should be just sufficient enough to enable ratio of residence time to diffusion time scale to be equal or greater than 1. Thus, we observed that the length of secondary lamella showed weak dependence on weight of fishes. However, the primary lamella being a structural element that acts as a rack for parallel stacking of secondary lamellae demonstrated a strong dependence on the weight of the fishes. This may be to maximize the number of accommodating secondary lamella which is the respiratory functional unit of fish gills. An increase in the mass of fishes is related to their energy requirements which can be fulfilled by proportionate increase in the surface area of fish gills. The increase in number of secondary lamella eventually increases the surface area available for gas/solute exchange. Therefore, we observed that surface area of gills and number of secondary lamella increases with an increase in fish body mass (Figure 3B). The surface area of fish gills scale linearly with weight of fishes. Thus, the length of the primary lamella demonstrates strong dependence on weight of fishes (Figure 3A). Therefore, the arrangement of primary and secondary lamellae may play a vital role in a head of fish body with different body masses.

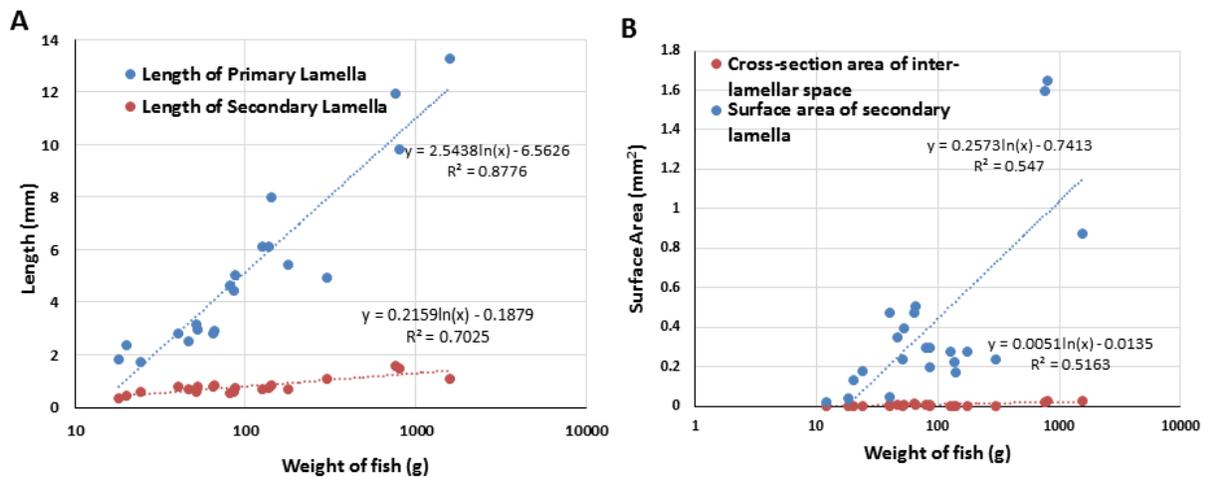

Figure 3 Graph showing the variation of A) length of primary and secondary lamella with weight of fishes on logarithmic scale b) surface area of secondary lamellae and cross-section area of inter-lamellar space with weight of fishes on logarithmic scale.

Further, the gill's primary and secondary lamella are arranged in a compact manner to enable large surface area per unit volume occupied by gills. The fractal dimensions calculated for secondary lamella shows that it is greater than 2 for most of the fishes (Figure 4). It is reported in literature that higher fractal dimensions indicates greater compactness of the structure. These fractal dimensions allows the gills to be compact and efficient increase surface area for effective gas exchange. In the absence of data for primary lamella of fish gills, we could not calculate their fractal dimensions. However, we believe that to ensure their compactness in the fish body, fractal dimensions for primary lamella of fish gills must be above 2. This arrangement of primary and secondary lamella ensures compactness of gills in fishes.

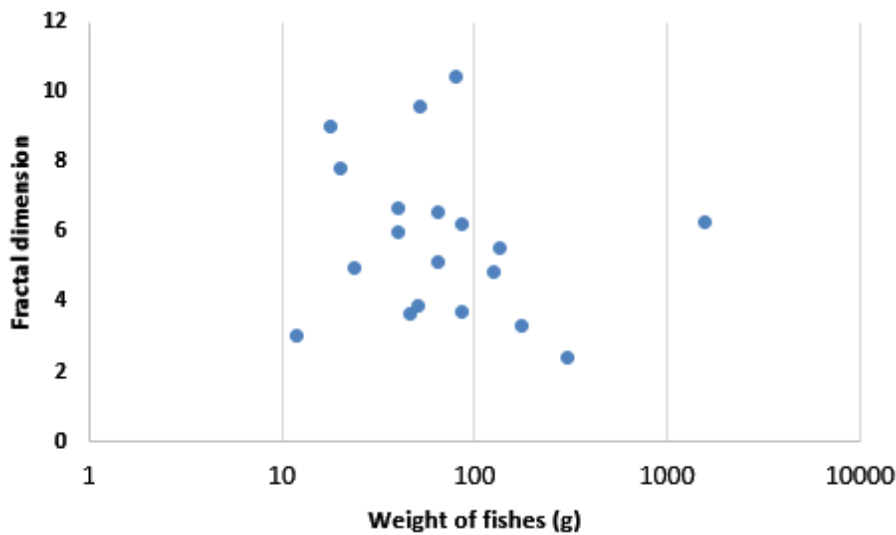

Figure 4 shows the variation of fractal dimensions of secondary lamella of fish gills with weight of different fishes

In order to develop a device for gas/solute exchange, mimicking the multi-scale architecture of fishes will help in developing micro/nanofluidic devices for mass transfer applications. However, to develop an understanding that can facilitate in translating the concept of mass transfer in fishes for developing scalable micro/nanofluidic devices, independent of the size of fishes will be highly desirable. Further, considering the multiple parameters of fish gills affecting the mass transfer, it is challenging to borrow dimensions from a fish to develop mass transfer device as per our requirement when the above parameter are closely coupled with other parameters. Therefore, having a few dimensionless parameters and knowing their degree of significance in mass transfer will assist in easy and efficient design of user defined size of mass transfer devices. Hence, few dimensional less parameters (parametric ratios) have been defined in this work. The rationale behind selecting these dimensionless parameters is to facilitate easy understanding of the effects of different parameters in their ability to affect the gas exchange process through convection-diffusion phenomenon. These dimensionless parameters summarizes majority of the structural and functional parameters in a four independent dimensionless numbers which governs the gas exchange through convection diffusion phenomena. Thus, the study of these parameters will help us in characterizing gas/solute exchange capabilities of any device where gas/solute exchange happens in two flowing fluid system separated by a membrane.

Although, the spatial-arrangement of primary and secondary lamella defines the architecture of gills and forms a basis for design and fabrication of compact biomimetic fish gill structure, they hardly gave any idea about the correlated factors affecting the solute/gas exchange at the site of secondary lamella. Therefore, the parametric ratios for gills of 24 different fishes were evaluated from their morphometric properties and compared. All the parametric ratios plotted against the weight of fishes on a logarithmic scale. It was observed that these parametric ratios remain nearly conserved in all 24 fishes taken for evaluation although their body mass varied from 12g to 1550g as shown in Figure 5. The thickness of membrane to water channel width ratio should be as small as possible to minimize the diffusion barrier length for solute/gases. The calculated data indeed suggested that this ratio for 24 different fishes lied in the range of 0.135 to 0.848 with an average of 0.372 ± 0.189, much less than 1. Further, the width of blood channel should neither be too less leading to high blood pumping pressure due to an increased blood flow resistance nor too high to leading to the condition of washout without proper oxygenation. Therefore, the ratio of width of blood channel to water channel lied between 0.071 and 0.449 with mean of 0.206 ± 0.095 in all 24 fishes. The width of the blood channel appear to be less in comparison with the width of the membrane

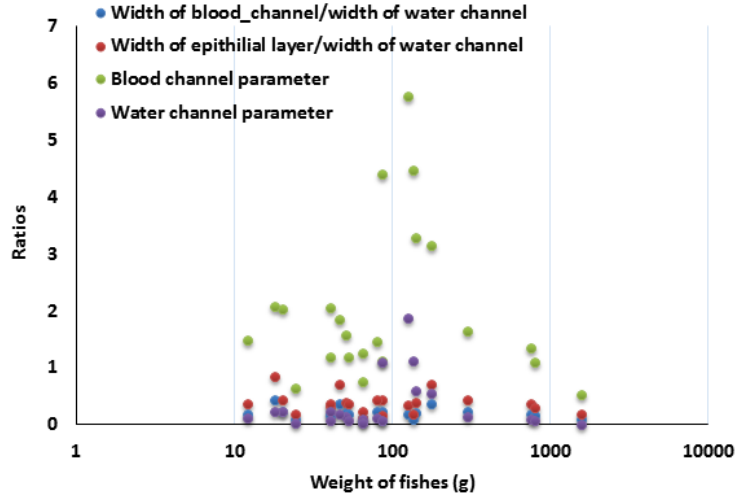

Figure 5 Graph showing the variation of different dimension less quantities (Parametric ratios) with weight of fishes on logarithmic scale.

Further, the blood and water parametric ratio defines the mixing parameter in blood and water channels, respectively. The mixing parameter is a dimensionless number which defines dominance of diffusion phenomenon over convection phenomenon in a laminar flow. The blood parametric ratio existed between 0.535 and 5.778 with mean value 2.11±1.41. The mean value is above 1 indicating the dominance of convective flow over diffusive flow. Such behavior leads to the continuous wash out of oxygen which enters in the blood stream, thereby maintaining a continuous concentration gradient between the blood and the water stream. The kinetics of capture of oxygen by hemoglobin of red blood cells defines the rate of oxygen capture and its removal by convective flow creates oxygen deficient center in the flowing blood stream. The minimum of blood parameter ratio obtained from the above data is 0.535. The large variation in the blood parameter is due to the assumptions on the dimensions of blood channels. When water parametric ratio was considered, it was found that this parameter in different fishes lie between 0.016 and 1.88 with mean of 0.323 ± 0.459. This clearly illustrates the dominance of diffusion phenomena over the convective transport phenomena in the water channel. Such behavior is desirable for fishes because it indicate that maximum of oxygen from given volume of water diffuses out of water stream and enters in blood stream before the water leave the inter-lamellar space. This condition ensures minimum washout of oxygen from the water stream while passing through inter-lamellar space. Although the mean value of water parametric ratio is below 1, the values greater than 1 just indicate the underutilization of gills for oxygen transport.

From Figure 5, it was observed that the data of 3 out of 4 different parametric ratios for different fishes showed minimum variation as compared to the weight of fishes. Majority of ratios lie within the window of 1±1. The blood parameter was found to be greater than 1, indicating the dominance of convective process over the diffusion process. This take care of the continuous maintenance of oxygen deficient region in the blood for continued diffusion of oxygen from interlamellar space to blood channel. The value of these parametric ratios do not vary much with the weight of fishes indicating their evolutionary conservation in fishes to maximize it efficiency of gas/solute exchange. Further, the architectural design of primary and secondary lamella is also conserved across different fishes to aid in enhance mass transport. Therefore, the above parametric ratios and arrangement of primary and secondary lamella can be used as engineering design parameters to develop micro/nanodevices for gas and solute exchange applications. However, these data do not speak about the nature of variation of different parametric ratios with the efficiency of oxygen exchange. Therefore, we have studied the gas exchange in a secondary lamella during convective–diffusion phenomena to evaluate the nature of these parametric ratios through computational method.

The concentration profile results of oxygen in different layers a) water channel b) thin epithelial layer and c) blood channel was shown in Figure 6B. From the concentration profile, it was evident that there always exist a concentration gradient at an entry of blood channel and exit of water channel, even at steady state condition, providing a continuous diffusion of oxygen from water channel to blood channel via thin epithelial barrier. Further, when oxygen concentration at an exit of blood stream was evaluated against each parametric ratio defined previously, each parameter plays a distinctive role in oxygen concentration flux at the exit of blood channel (Figure 6A). It was observed that thickness of epithelial barrier should be as thin as possible as compared to the water channel to maximize the exchange of gas across the epithelial barrier. Moreover, the width of the blood channel also needs to be small as compared to water channel. The oxygen concentration at the exit of blood channel decreases rapidly with increase in the ratios of width of thin epithelial membrane to the width of the water channel and width of the blood channel to the width of the water channel, respectively (Figure 6A). However, the width of the blood channel need to be of an optimal size to avoid high pumping pressure to overcome increased blood flow resistance and assist in maximum diffusion of oxygen in the blood stream before oxygenated blood comes out of blood channel. Further, the blood parametric ratio suggests that above a critical ratio of 0.33, the effect of increase in ratio hardly affect the oxygen concentration at the exit of blood channel. Below the critical ratio, the diffusion phenomenon dominates the convection phenomena, thereby leading to low oxygen concentration in the blood at the exit. However, at a higher ratio, blood saturated with oxygen washes out of secondary lamella. This suggests that higher ratios are preferable for better gas exchange. The water parametric ratio has negligible effect on the oxygenation of blood within the range of ratios considered in the study. However, at a much higher ratio, the decrease in concentration of oxygen in blood clearly indicated the case of washout. It was observed that all the curves pertaining to different parametric ratios intersect at ratio of 0.33. This ratio marks the optimal ratio of different parameters to enable maximum exchange of gas from water stream to blood stream and is shown as a yellow star in the Figure 6A .

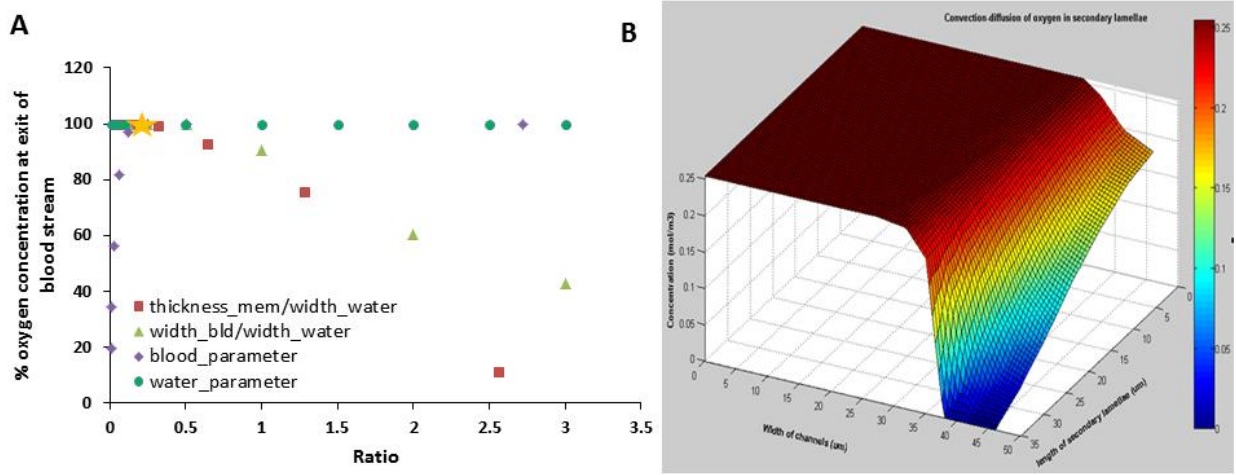

Figure 6 A) Graph showing the effect of different dimensional less parameters (ratios) on % of oxygen concentration at an exit of blood channel a) Concentration profile of oxygen in different in blood and water channels.

On comparison of results obtained from theoretical and computational model, it was found that overall results show good agreement with each other (Figure 7). Although, the three parametric ratios (ratio of thickness of membrane to width of water channel, ratio of width of blood channel to the width of water channel and water parametric ratios) predicted by computational model is in congruence with theoretical results as shown in figure, the blood parametric ratio calculated through computational model demonstrated significant difference from theoretical results.

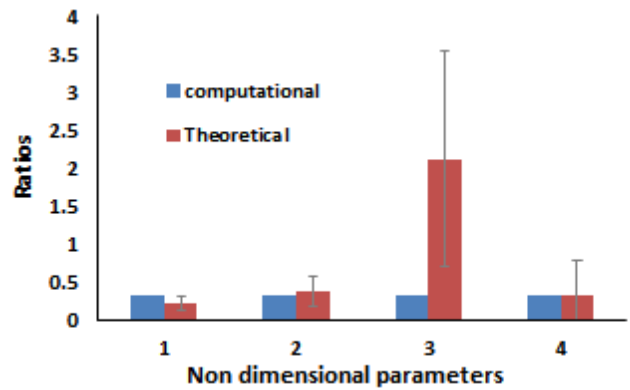

Figure 7 Graph showing the comparison of parametric ratios evaluated from computational method and theoretical analysis from data available in literature

This might be due to fixed velocity assumed in blood channel for all the fishes in the theoretical analysis. The velocity of blood in the secondary lamella of fish varies with species of fishes, activity of fishes and water medium in which they thrive. Therefore, fishes manipulate the blood flow velocity in accordance with the width of inter lamellar space to maximize the sustained diffusion of oxygen from water to blood of fishes. Eventually, our study suggested few parametric ratios and structural parameters that are responsible for higher performance efficiency of gills in mass transfer.

## 5. Conclusion

Fish gills are one of the excellent gas/solute exchange structure available in nature. Such performances are attributed to their extraordinary multi-scale hierarchical structures. Therefore, bio-mimicking these structures at different length scales to form a gas/solute exchange device require detailed study of their structural properties and how they guide the transport properties in fish gills. Further, there is also a need to formulate few dimensionless parameters which can be used as design parameters for development of gas/solute separation device. Hence, we have established four parametric ratios which affect the convection-diffusion phenomenon in fishes and studied their role in gas exchange through modelling and simulation. It was observed that these parametric ratios tend to have an optimal value around 0.33 for the maximization of gas exchange cross secondary lamella. Further, our theoretical analysis on 24 different fishes supported our argument about very limited variability of above parametric ratios with weight of fishes. Thus, structural design of biomimetic gas/solute exchange devices should incorporate these parametric ratios as design parameter. Apart from these parametric ratios, the role of the factors like dimensions of primary and secondary lamella, surface area of fish gills and inter-lamellar distance have been studied in detail. These studies can form a backbone for the biomimetic development of devices for gas/solute exchange.

## Acknowledgement

This research effort was supported partially by Suman Mashruwala Microengineering Laboratory (www.me.iitb.ac.in/~mems) sponsored by IIT Bombay alumnus Mr. Raj Mashruwala and IITB-Monash Research Academy ( A joint venture between IIT Bombay, India and Monash University, Clayton, Melbourne, Australia). Few results in this piece of work has been published as a conference proceedings in international conference, SPIE/NDE, (Las Vegas, USA), 9797-40, 2016.